\documentclass[prb,twocolumn,groupedaddress,shownopacs,amssymb]{revtex4-2}
\usepackage{graphicx,psfrag,amsmath,amssymb}

\usepackage[usenames, dvipsnames]{color}

\usepackage{hyperref} 
\hypersetup{
    colorlinks=true,
    linkcolor=Blue,
    citecolor=Blue,
    filecolor=Blue,      
    urlcolor=Blue,
    pdftitle={pair size},
    }

\begin{document}

\title{Pair size and quantum geometry in a multiband Hubbard model}

\author{M. Iskin}
\affiliation{
Department of Physics, Ko\c{c} University, Rumelifeneri Yolu, 
34450 Sar\i yer, Istanbul, T\"urkiye
}

\date{\today}

\begin{abstract}

We study the size of two-body bound states and Cooper pairs within a 
multiband Hubbard model that features time-reversal symmetry and uniform 
pairing on a generic lattice. Our analysis involves (i) an exact 
calculation of the localization tensor to determine the size of 
lowest-lying two-body bound state in vacuum, and (ii) an evaluation 
of the analogous tensor to estimate the average size of Cooper pairs within 
the mean-field BCS-BEC crossover theory at zero temperature. 
Beyond the conventional intraband 
contribution that depends on Bloch bands, we show that pair size 
also has a geometric contribution governed by the quantum-metric tensor 
of the Bloch states and their band-resolved quantum-metric tensors. 
As a concrete example, we investigate the pyrochlore-Hubbard model 
numerically and demonstrate that, while the pair size diverges in the 
weakly interacting BCS regime of dispersive bands, it remains finite 
and relatively small in the flat-band regime, even for infinitesimal 
interaction, perfectly matching the exact two-body result in the dilute limit.

\end{abstract}

\maketitle

\section{Introduction}
\label{sec:intro}

The quantum geometry of a Bloch band is characterized by the quantum-metric 
tensor, which corresponds to the real part of the quantum-geometric tensor 
and measures the so-called quantum distance between two nearby Bloch 
states~\cite{resta11, torma23}. Recent studies on multiband Hubbard models 
have revealed that this band invariant plays a crucial role in describing 
virtual interband processes that influence key observables in superconductivity. 
These include, but are not limited to, the superfluid weight, superfluid 
density, critical transition temperature, low-energy collective excitations, 
and the Ginzburg-Landau (GL) coherence 
length~\cite{torma22, huhtinen22, herzog22, iskin24, daido23, 
iskin23, chen23, verma24, jiang23, jiang24, kitamura24, iskin24c,
han24, hu24}. 
The quantum geometry inherent in multiband systems modifies these 
observables, demonstrating that the quantum metric is not merely a 
mathematical artifact but a central quantity in determining the physical 
behavior of superconductors, particularly in flat-band systems where 
geometric effects are more pronounced. It is important to emphasize that 
these observables are not independent but are instead directly linked 
through the effective-mass tensor of Cooper pairs~\cite{iskin18b, iskin24}. 
This connection is also rooted in the effective-mass tensor of the 
lowest-lying two-body bound states in vacuum~\cite{torma22, iskin21}. 
For this reason, the geometric nature of superconducting pairing manifests 
itself across multiple physical quantities, showing that the quantum metric 
is an essential component of the underlying physics in multiband 
superconductors. 

In this study, we investigate the role of quantum geometry in determining 
the size of two-body bound states and Cooper pairs within a generic 
multiband Hubbard model that exhibits time-reversal symmetry and 
uniform pairing. We analyze the localization tensors for the 
lowest-lying two-body bound state in vacuum and the average size of 
Cooper pairs within the mean-field BCS-BEC crossover theory at zero 
temperature. Our main result highlights a stark contrast between the 
BCS regime of dispersive bands and the flat-band regime: while the 
pair size diverges in the weakly interacting BCS regime, it remains 
finite and relatively small in the flat-band regime, even for infinitesimal
interactions, where it is governed entirely by the quantum metric.
It is pleasing to note that this result is consistent with the modern 
theory of insulating states, where the localization tensor diverges 
in the thermodynamic limit for metals, while it remains finite for
insulators~\cite{resta11}. 
We also emphasize that pair size and coherence length, though related, 
are distinct physical quantities with different quantum-geometric origins. 
The coherence length is linked to the motion of Cooper pairs via 
the inverse effective-mass tensor~\cite{iskin23, chen23, iskin24c}, 
whereas the pair size does not have a direct relationship with the 
effective mass or center-of-mass momentum of the pair, distinguishing 
it from previous studies in this context. Away from the flat-band 
regime, the pair size mirrors the zero-temperature coherence length~\cite{iskin24c}, scaling inversely with the order parameter 
across much of the parameter space. However, in stark contrast to 
the finite and relatively small pair size in the dilute flat-band 
regime, the coherence length has recently been shown to diverge in 
this limit~\cite{iskin24c}.

The remaining text is organized as follows. In Sec.~\ref{sec:Hubbard}, 
we introduce the multiband Hubbard model in reciprocal space.
In Sec.~\ref{sec:twobody}, we derive the size of two-body bound states
in vacuum through an exact calculation of the localization tensor.
In Sec.~\ref{sec:BCS}, we estimate the average size of Cooper pairs
within the mean-field BCS-BEC crossover theory at zero temperature.
In Sec.~\ref{sec:numerics}, we present the numerical calculations for 
the pyrochlore-Hubbard model. The paper concludes with a summary 
in Sec.~\ref{sec:conc}, and an alternative but failed approach to 
the size of Cooper pairs is discussed in the Appendix.

\section{Multiband Hubbard model}
\label{sec:Hubbard}

The multiband Hubbard model 
$
\mathcal{H} = \sum_\sigma \mathcal{H}_\sigma + \mathcal{H}_{\uparrow\downarrow}
$
consists of two parts. The non-interacting part
$
\mathcal{H}_\sigma = -\sum_{\mathrm{i} \mathrm{i}' S S'} 
t_{\mathrm{i} S; \mathrm{i}' S'}^\sigma 
c_{S \mathrm{i} \sigma}^\dagger c_{S' \mathrm{i}' \sigma}
$
describes the hopping processes between lattice sites, where the operator
$c_{S i \sigma}^\dagger$ creates a spin-$\sigma$ fermion on the sublattice site 
$S$ in the $\mathrm{i}$th unit cell, and $t_{\mathrm{i} S; \mathrm{i}' S'}^\sigma$
is the hopping parameter between site $S'$ in unit cell $\mathrm{i}'$ and 
site $S$ in unit cell $\mathrm{i}$. 
To reexpress $\mathcal{H}_\sigma$ in reciprocal space,
we introduce a canonical transformation,
$
c_{S \mathrm{i} \sigma}^\dagger = \frac{1}{\sqrt{N_c}} \sum_\mathbf{k} 
e^{-i \mathbf{k} \cdot \mathsf{r}_{\mathrm{i} S}} c_{S \mathbf{k} \sigma}^\dagger,
$
where $N_c$ is the number of unit cells in the system, 
$
\mathbf{k} = (k_x, k_y, k_z)
$ 
is the crystal momentum (in units of $\hbar \to 1$) within the first 
Brillouin zone (BZ), and $\mathsf{r}_{\mathrm{i} S}$ is the position of site 
$S \in \mathrm{i}$. The momentum summation satisfies 
$
\sum_{\mathbf{k} \in \mathrm{BZ}} 1 = N_c,
$ 
and when the number of sublattice sites within a unit cell is $N_b$, 
the total number of lattice sites in the system is $N = N_b N_c$.
This transformation leads to the Bloch Hamiltonian in the sublattice 
basis,
\begin{align}
\mathcal{H}_\sigma = \sum_{S S' \mathbf{k}} h_{SS'\mathbf{k}}^\sigma
c_{S \mathbf{k} \sigma}^\dagger c_{S' \mathbf{k} \sigma},
\end{align}
where the matrix elements
$
h_{SS'\mathbf{k}}^\sigma = -\frac{1}{N_c}\sum_{\mathrm{i} \mathrm{i}'} 
t_{\mathrm{i}S; \mathrm{i}'S'}^\sigma
e^{\mathrm{i} \mathbf{k} \cdot 
(\mathsf{r}_{\mathrm{i} S} - \mathsf{r}_{\mathrm{i'} S'})}
$
determine the single-particle spectrum via the eigenvalue relation
\begin{align}
\sum_{S'} h_{SS'\mathbf{k}}^\sigma n_{S' \mathbf{k} \sigma} 
= \varepsilon_{n\mathbf{k} \sigma} n_{S \mathbf{k} \sigma}.
\end{align}
Here, $\varepsilon_{n\mathbf{k} \sigma}$ represents the energy of the $n$th
Bloch band, and $n_{S \mathbf{k} \sigma}$ is the periodic part of the
corresponding Bloch wave function, as discussed further below. Finally,
applying the basis transformation
$
c_{S \mathbf{k} \sigma}^\dagger = \sum_n n_{S \mathbf{k} \sigma}^* 
c_{n \mathbf{k} \sigma}^\dagger,
$
we recast the Hamiltonian in the band basis as
$
\mathcal{H}_\sigma = \sum_{n \mathbf{k}} 
\varepsilon_{n\mathbf{k}\sigma}
c_{n \mathbf{k} \sigma}^\dagger c_{n \mathbf{k} \sigma}.
$
On the other hand, the interacting part
$
\mathcal{H}_{\uparrow\downarrow} = - U \sum_{\mathrm{i} S} 
c_{S \mathrm{i} \uparrow}^\dagger c_{S \mathrm{i} \downarrow}^\dagger 
c_{S \mathrm{i} \downarrow} c_{S \mathrm{i} \uparrow}
$
with $U \ge 0$ describes the on-site density-density attraction between
spin-$\uparrow$ and spin-$\downarrow$ fermions. In reciprocal space, 
we reexpress it as
\begin{align}
\mathcal{H}_{\uparrow \downarrow} = 
- \frac{U}{N_c}\sum_{S \mathbf{k} \mathbf{k'} \mathbf{q}}
c_{S, \mathbf{k} + \frac{\mathbf{q}}{2} \uparrow}^\dagger 
c_{S, -\mathbf{k} + \frac{\mathbf{q}}{2}, \downarrow}^\dagger 
c_{S, -\mathbf{k'} + \frac{\mathbf{q}}{2}, \downarrow} 
c_{S, \mathbf{k'} + \frac{\mathbf{q}}{2}, \uparrow}.
\end{align}
Next, we take advantage of the explicit conservation of total momentum 
in the $\mathbf{k}$-space formulation and present an exact solution for 
the two-body problem.

\section{Exact two-body problem}
\label{sec:twobody}

To solve for the spin-singlet bound states of a pair of spin-$\uparrow$ 
and spin-$\downarrow$ fermions with center-of-mass momentum $\mathbf{q}$, 
we adopt the ansatz state~\cite{iskin21}
\begin{align}
|\Psi_\mathbf{q} \rangle = \sum_{n m \mathbf{k}} 
\alpha_{nm\mathbf{k}}^\mathbf{q} 
c_{n, \mathbf{k}+\frac{\mathbf{q}}{2}, \uparrow}^\dagger 
c_{m,-\mathbf{k}+\frac{\mathbf{q}}{2},\downarrow}^\dagger 
| 0 \rangle,
\label{eqn:ansatz}
\end{align}
where $|0 \rangle$ is the vacuum state, and $\alpha_{nm\mathbf{k}}^\mathbf{q}$ 
are the variational parameters. The requirement
$
\alpha_{nm\mathbf{k}}^\mathbf{q} = \alpha_{mn,-\mathbf{k}}^\mathbf{q}
$
ensures that $|\Psi_\mathbf{q} \rangle$ is antisymmetric under the exchange 
of fermions, which is necessary and sufficient for the spin-singlet state.
The normalization condition 
$
\langle \Psi_\mathbf{q} |\Psi_\mathbf{q} \rangle = 1
$
leads to the constraint
$
\sum_{n m \mathbf{k}} |\alpha_{nm\mathbf{k}}^\mathbf{q}|^2 = 1.
$
For any given $\mathbf{q}$, minimization of the expectation value 
$
\langle \Psi_\mathbf{q} | \mathcal{H} - E_\mathbf{q} |\Psi_\mathbf{q} \rangle
$
with respect to $\alpha_{nm\mathbf{k}}^\mathbf{q}$, where $E_\mathbf{q}$ 
represents the energy of the allowed two-body states, leads to a set 
of linear equations~\cite{iskin21}
\begin{align}
\alpha_{n m \mathbf{k}}^\mathbf{q} = 
\frac{
\frac{U}{N_c}\sum_{S}\beta_{S \mathbf{q}}
{n}_{S, \mathbf{k}+\frac{\mathbf{q}}{2}, \uparrow}^*
{m}_{S, -\mathbf{k}+\frac{\mathbf{q}}{2}, \downarrow}^*
}
{\varepsilon_{n, \mathbf{k} +\frac{\mathbf{q}}{2}, \uparrow}
+ \varepsilon_{m, -\mathbf{k}+\frac{\mathbf{q}}{2},\downarrow} 
- E_\mathbf{q}}.
\label{eqn:alphanmk}
\end{align}
Here, the dressed variational parameters
$
\beta_{S \mathbf{q}} 
= \sum_{n m \mathbf{k}}
\alpha_{n m \mathbf{k}}^\mathbf{q}
{n}_{S, \mathbf{k}+\frac{\mathbf{q}}{2}, \uparrow}
{m}_{S, -\mathbf{k}+\frac{\mathbf{q}}{2}, \downarrow}
$
characterize the physical properties of the two-body bound states, 
and $E_\mathbf{q}$ are given by the eigenvalues of an 
$
N_b^2 N_c \times N_b^2 N_c
$ 
matrix through Eq.~(\ref{eqn:alphanmk}).
Next, we discuss the two-body wave function and its localization tensor 
from which the pair size follows.

\subsection{Localization tensor for the bound states}
\label{sec:tensor}

The wave function for the resultant two-body bound states is determined by
$
\Psi_\mathbf{q} (\mathsf{r}_1,\mathsf{r}_2) = 
\langle \mathsf{r}_1 \mathsf{r}_2 | \Psi_\mathbf{q} \rangle,
$
where 
$
\mathsf{r}_1 = (x_1, y_1, z_1)
$
and
$
\mathsf{r}_2 = (x_2, y_2, z_2)
$
are the Cartesian components. Here, we express the Bloch wave function 
for a spin-$\sigma$ particle as
$
\phi_{n \mathbf{k} \sigma} (\mathsf{r}) = 
\langle \mathsf{r} | n \mathbf{k} \sigma \rangle = 
\frac{e^{i \mathbf{k} \cdot \mathsf{r}}}{\sqrt{N_c}} 
n_{\mathbf{k} \sigma} (\mathsf{r}),
$
where
$
| n \mathbf{k} \sigma \rangle = c_{n \mathbf{k} \sigma}^\dagger |0\rangle
$
and 
$
n_{\mathbf{k} \sigma} (\mathsf{r})
$
represents the periodic part of the wave function. This leads to
$
\Psi_\mathbf{q} (\mathsf{r}_1,\mathsf{r}_2) =
\frac{e^{i \mathbf{q} \cdot \mathbf{R}}}{N_c}
\sum_{nm \mathbf{k}} e^{i \mathbf{k} \cdot \mathbf{r}} 
\alpha_{n m \mathbf{k}}^\mathbf{q}
{n}_{\mathbf{k}+\frac{\mathbf{q}}{2}, \uparrow} (\mathsf{r}_1)
{m}_{-\mathbf{k}+\frac{\mathbf{q}}{2}, \downarrow} (\mathsf{r}_2),
$
where 
$
\mathbf{R} = (\mathsf{r}_1+\mathsf{r}_2)/2
$
is the center-of-mass position of the pair, and 
$
\mathbf{r} = \mathsf{r}_1-\mathsf{r}_2
$
is the relative position of its constituents. Note that 
$
\Psi_\mathbf{q} (\mathsf{r}_1, \mathsf{r}_2) =
\Psi_\mathbf{q} (\mathsf{r}_2, \mathsf{r}_1)
$
upon spin exchange by construction when $\uparrow \leftrightarrow \downarrow$. 
For the multiband Hubbard model of interest in this paper, we substitute 
$
\mathsf{r}_1 \to \mathsf{r}_{\mathrm{i} S}
$ 
and 
$
\mathsf{r}_2 \to \mathsf{r}_{\mathrm{i'} S'}, 
$
leading to
$
\Psi_\mathbf{q} (\mathsf{r}_{\mathrm{i} S},\mathsf{r}_{\mathrm{i'} S'}) =
e^{i \mathbf{q} \cdot (\mathsf{r}_{\mathrm{i} S} +\mathsf{r}_{\mathrm{i'} S'})/2}
\psi_{SS'}^\mathbf{q}(\mathbf{\bar{r}}),
$
where
$
\mathbf{\bar{r}} = \mathsf{r}_{\mathrm{i} S} - \mathsf{r}_{\mathrm{i'} S'}
$
is the relative position. The latter function is defined as
\begin{align}
\psi_{SS'}^\mathbf{q}(\mathbf{\bar{r}}) =
\frac{1}{N_c} \sum_{nm \mathbf{k}} 
e^{i \mathbf{k} \cdot \mathbf{\bar{r}}} 
\alpha_{n m \mathbf{k}}^\mathbf{q}
{n}_{S, \mathbf{k}+\frac{\mathbf{q}}{2}, \uparrow}
{m}_{S', -\mathbf{k}+\frac{\mathbf{q}}{2}, \downarrow},
\end{align}
with the constraint
$
\sum_{\mathrm{i} S \mathrm{i'} S'} 
|\psi_{SS'}^\mathbf{q}(\mathbf{\bar{r}})|^2 
= \sum_{nm\mathbf{k}} |\alpha_{nm\mathbf{k}}^\mathbf{q}|^2 = 1.
$
This normalization condition follows simply from the orthonormalization 
conditions
$
\frac{1}{N_c} \sum_{\mathrm{i}} 
e^{i (\mathbf{k} - \mathbf{k'}) \cdot \mathsf{r}_{\mathrm{i} S}}
= \delta_{\mathbf{k} \mathbf{k'}}
$
and
$
\sum_S n_{S \mathbf{k} \sigma}^* m_{S \mathbf{k} \sigma}
= \langle n_{\mathbf{k} \sigma} | m_{\mathbf{k} \sigma} \rangle
= \delta_{nm},
$
where $\delta_{ij}$ is the Kronecker-delta. 

For the simplicity of the presentation, we denote the Cartesian components 
of the relative position as 
$
\mathbf{r} = (r_x, r_y, r_z)
$ 
where $r_x = x_1 - x_2$, $r_y = y_1 - y_2$, and $r_z = z_1 - z_2$. 
It is easy to see that
$
\langle \Psi_\mathbf{q}| r_i | \Psi_\mathbf{q} \rangle 
= \sum_{\mathsf{r}_1 \mathsf{r}_2} r_i|\Psi_\mathbf{q} (\mathsf{r}_1,\mathsf{r}_2)|^2 = 0,
$
since changing the dummy summation indices 
$
\mathsf{r}_1 \leftrightarrow \mathsf{r}_2
$ 
changes the overall sign of the summand. Then, to extract the pair size 
from the variance of the relative position, we introduce the so-called 
localization tensor~\cite{resta11}, whose real and symmetric matrix elements 
are given by
$
(\xi_{2b}^2)_{ij} 
= \langle \Psi_\mathbf{q}| r_i r_j | \Psi_\mathbf{q} \rangle 
= \sum_{\mathsf{r}_1 \mathsf{r}_2} r_i r_j 
|\Psi_\mathbf{q} (\mathsf{r}_1, \mathsf{r}_2)|^2.
$
For the multiband Hubbard model of interest in this paper, 
it can be written as
\begin{align}
(\xi_{2b}^2)_{ij} = \sum_{\mathrm{i} S \mathrm{i'} S'}
\bar{r}_i \bar{r}_j |\psi_{SS'}^\mathbf{q}(\mathbf{\bar{r}})|^2,
\label{eqn:loctensor}
\end{align}
where $\bar{r}_i$ is the $i$-component of $\mathbf{\bar{r}}$.
Then, we substitute 
$
r_i e^{i \mathbf{k} \cdot \mathbf{r}} 
= -i \partial_i (e^{i \mathbf{k} \cdot \mathbf{r}})
$
and
$
r_j e^{-i \mathbf{k'} \cdot \mathbf{r}} 
= i \partial_{j'} (e^{-i \mathbf{k'} \cdot \mathbf{r}}),
$
where $\partial_i \equiv \frac{\partial}{\partial k_i}$ and 
$\partial_{j'} \equiv \frac{\partial}{\partial k_j'}$,
and use Green's theorem for periodic functions~\cite{ashcroft}, i.e.,
integration by parts. This shifts the partial derivatives from the 
exponential factors to the Bloch factors. Summing over the unit cells 
through the orthonormalization condition for the exponential factors, 
we eventually find
\begin{align}
(\xi_{2b}^2)_{ij} = 
\sum_{nm n'm' SS' \mathbf{k}} &
\partial_i (\alpha_{n m \mathbf{k}}^\mathbf{q}
{n}_{S, \mathbf{k}+\frac{\mathbf{q}}{2}, \uparrow}
{m}_{S', -\mathbf{k}+\frac{\mathbf{q}}{2}, \downarrow})
\nonumber \\ \times
\partial_j & (\alpha_{n' m' \mathbf{k}}^\mathbf{q}
{n'}_{S, \mathbf{k}+\frac{\mathbf{q}}{2}, \uparrow}
{m'}_{S', -\mathbf{k}+\frac{\mathbf{q}}{2}, \downarrow})^*.
\label{eqn:xi_ij}
\end{align}
This expression can also be reexpressed in alternative forms
~\footnote{For instance, using the Green's theorem for periodic 
functions~\cite{ashcroft}, it can be written as
$
(\xi_{2b}^2)_{ij} = - \sum_{nmn'm' SS' \mathbf{k}}
\alpha_{n m \mathbf{k}}^\mathbf{q}
{n}_{S, \mathbf{k}+\frac{\mathbf{q}}{2}, \uparrow}
{m}_{S', -\mathbf{k}+\frac{\mathbf{q}}{2}, \downarrow}
$
$
\partial_i \partial_j (\alpha_{n' m' \mathbf{k}}^\mathbf{q}
{n'}_{S, \mathbf{k}+\frac{\mathbf{q}}{2}, \uparrow}
{m'}_{S', -\mathbf{k}+\frac{\mathbf{q}}{2}, \downarrow})^*
$
}. 
After performing the derivatives, it leads to a relatively complicated
expression
\begin{widetext}
\begin{align}
(\xi_{2b}^2)_{ij} &= 
\sum_{nm\mathbf{k}} 
\partial_i \alpha_{n m \mathbf{k}}^\mathbf{q}
\partial_j (\alpha_{n m \mathbf{k}}^\mathbf{q})^*
+ 
\sum_{nm\mathbf{k}} 
\partial_i \alpha_{n m \mathbf{k}}^\mathbf{q} (\alpha_{n' m' \mathbf{k}}^\mathbf{q})^*
\Big( 
\langle \partial_j {n'}_{\mathbf{k}+\frac{\mathbf{q}}{2}, \uparrow}|
n_{\mathbf{k}+\frac{\mathbf{q}}{2}, \uparrow} \rangle \delta_{mm'}
+
\langle \partial_j {m'}_{-\mathbf{k}+\frac{\mathbf{q}}{2}, \downarrow}|
m_{-\mathbf{k}+\frac{\mathbf{q}}{2}, \downarrow} \rangle \delta_{nn'}
\Big)
\nonumber \\
& +
\sum_{nm n'm'\mathbf{k}} 
\alpha_{n m \mathbf{k}}^\mathbf{q} \partial_j (\alpha_{n' m' \mathbf{k}}^\mathbf{q})^*
\Big( 
\langle {n'}_{\mathbf{k}+\frac{\mathbf{q}}{2}, \uparrow}|
\partial_i n_{\mathbf{k}+\frac{\mathbf{q}}{2}, \uparrow} \rangle \delta_{mm'}
+
\langle {m'}_{-\mathbf{k}+\frac{\mathbf{q}}{2}, \downarrow}|
\partial_i m_{-\mathbf{k}+\frac{\mathbf{q}}{2}, \downarrow} \rangle \delta_{nn'}
\Big)
\nonumber \\
& +
\sum_{nm n'm'\mathbf{k}} 
\alpha_{n m \mathbf{k}}^\mathbf{q} (\alpha_{n' m' \mathbf{k}}^\mathbf{q})^*
\Big( 
\langle \partial_j {n'}_{\mathbf{k}+\frac{\mathbf{q}}{2}, \uparrow}|
\partial_i n_{\mathbf{k}+\frac{\mathbf{q}}{2}, \uparrow} \rangle \delta_{mm'}
+
\langle \partial_j {m'}_{-\mathbf{k}+\frac{\mathbf{q}}{2}, \downarrow}|
\partial_i m_{-\mathbf{k}+\frac{\mathbf{q}}{2}, \downarrow} \rangle \delta_{nn'}
\nonumber \\
& \quad\quad\quad\quad\quad +
\langle {n'}_{\mathbf{k}+\frac{\mathbf{q}}{2}, \uparrow}|
\partial_i n_{\mathbf{k}+\frac{\mathbf{q}}{2}, \uparrow} \rangle
\langle \partial_j {m'}_{-\mathbf{k}+\frac{\mathbf{q}}{2}, \downarrow}|
m_{-\mathbf{k}+\frac{\mathbf{q}}{2}, \downarrow} \rangle
+ 
\langle \partial_j {n'}_{\mathbf{k}+\frac{\mathbf{q}}{2}, \uparrow}|
n_{\mathbf{k}+\frac{\mathbf{q}}{2}, \uparrow} \rangle
\langle {m'}_{-\mathbf{k}+\frac{\mathbf{q}}{2}, \downarrow}|
\partial_i m_{-\mathbf{k}+\frac{\mathbf{q}}{2}, \downarrow} \rangle
\Big).
\label{eqn:xi_gen}
\end{align}
\end{widetext}
To make further analytical progress, we next consider a generic multiband 
Hubbard model that exhibits time-reversal symmetry and uniform pairing 
across the underlying sublattices within a unit cell, and focus on the 
lowest-lying bound state with $\mathbf{q} = \mathbf{0}$.

\subsection{Uniform-pairing condition}
\label{sec:ups}

In the remaining text, we assume that the Bloch Hamiltonian manifests 
time-reversal symmetry, where
$
h_{SS' \mathbf{k}}^\uparrow = (h_{SS',-\mathbf{k}}^\downarrow)^*,
$
which implies that
$
n_{S, -\mathbf{k}, \downarrow}^* = n_{S \mathbf{k} \uparrow} 
\equiv n_{S \mathbf{k}} 
$
for the Bloch factors and 
$
\varepsilon_{n, -\mathbf{k}, \downarrow} = \varepsilon_{n \mathbf{k} \uparrow} 
\equiv \varepsilon_{n \mathbf{k}}
$
for the Bloch bands. Furthermore, we assume that the so-called 
uniform-pairing condition, i.e., 
$
\beta_{S \mathbf{q}} \equiv \beta_{\mathbf{q}} 
$
for every sublattice site $S$ within a unit cell, is satisfied for the 
lowest-lying bound states $E_\mathbf{q}$ in the $\mathbf{q} \to \mathbf{0}$ 
limit. Under these assumptions, $E_\mathbf{q}$ can be Taylor expanded as
$
E_\mathbf{q} = E_b + \frac{1}{2}\sum_{ij} (M_{2b}^{-1})_{ij} q_i q_j + \cdots,
$
where $E_b = E_\mathbf{0}$ is the energy offset determined by
$
1 = \frac{U}{N} \sum_{n \mathbf{k}} 1/(2\varepsilon_{n \mathbf{k}} - E_b)
$
and the matrix elements $(M_{2b}^{-1})_{ij}$ constitute the inverse 
effective-mass tensor~\cite{iskin21}. 
In addition, Eq.~(\ref{eqn:alphanmk}) reduces to
$
\alpha_{n m \mathbf{k}}^\mathbf{0} = \alpha_{nn \mathbf{k}}^\mathbf{0} \delta_{nm} 
$
at $\mathbf{q} = \mathbf{0}$, where
$
\alpha_{nn \mathbf{k}}^\mathbf{0} = U \beta_\mathbf{0}/
[N_c (2 \varepsilon_{n \mathbf{k}} - E_b)],
$
and the normalization condition requires
$
\big( \frac{N_c}{U |\beta_\mathbf{0}|} \big)^2 = 
\sum_{n \mathbf{k}} \frac{1}{(2 \varepsilon_{n \mathbf{k}} - E_b)^2}.
$
Noting that 
$
\alpha_{nn \mathbf{k}}^\mathbf{0} (\alpha_{mm \mathbf{k}}^\mathbf{0})^*
$
is a real number and using
$
\langle \partial_i n_{\mathbf{k} \sigma} | m_{\mathbf{k} \sigma} \rangle
= -\langle n_{\mathbf{k} \sigma} | \partial_i m_{\mathbf{k} \sigma} \rangle
$
in Eq.~(\ref{eqn:xi_gen}), we can express the localization tensor as
$
(\xi_{2b}^2)_{ij} = (\xi_{2b}^2)^\mathrm{intra}_{ij} 
+ (\xi_{2b}^2)^\mathrm{inter}_{ij},
$
where 
$
(\xi_{2b}^2)^\mathrm{intra}_{ij} = \sum_{n \mathbf{k}} 
\partial_i \alpha_{nn \mathbf{k}}^\mathbf{0} \partial_j 
(\alpha_{nn \mathbf{k}}^\mathbf{0})^*
$
is the intraband contribution and
$
(\xi_{2b}^2)^\mathrm{inter}_{ij} = 
\sum_{n \mathbf{k}} |\alpha_{nn \mathbf{k}}^\mathbf{0}|^2 g_{ij}^{n \mathbf{k}}
- \sum_{n, m \ne n, \mathbf{k}} \alpha_{nn \mathbf{k}}^\mathbf{0} 
(\alpha_{mm \mathbf{k}}^\mathbf{0})^*
g_{ij}^{nm \mathbf{k}}
$
is the interband contribution.
Here,
$
g_{ij}^{n\mathbf{k}} = \sum_{m \ne n} g_{ij}^{nm\mathbf{k}}
$
is the quantum-metric tensor of the $n$th Bloch band, where
\begin{align}
\label{eqn:metric}
g_{ij}^{nm\mathbf{k}} = 2\mathrm{Re} \langle 
\partial_i n_\mathbf{k} | m_\mathbf{k} \rangle
\langle m_\mathbf{k} | \partial_j n_\mathbf{k} \rangle
\end{align}
is the so-called band-resolved quantum-metric tensor, with $\mathrm{Re}$ 
denoting the real part. Note that while
$
g_{ij}^{n\mathbf{k}} = g_{ji}^{n\mathbf{k}} 
$
is symmetric, 
$
g_{ij}^{nm\mathbf{k}} = g_{ji}^{mn\mathbf{k}}
$
is not. More explicitly, the alternative expressions 
\begin{align}
\label{eqn:2bintra}
(\xi_{2b}^2)^\mathrm{intra}_{ij} &= \frac{
4\sum_{n \mathbf{k}} \frac{\partial_i \varepsilon_{n\mathbf{k}}
\partial_j \varepsilon_{n\mathbf{k}}}{(2\varepsilon_{n \mathbf{k}} - E_b)^4}}
{\sum_{n \mathbf{k}} \frac{1}{(2\varepsilon_{n \mathbf{k}} - E_b)^2}}, \\
(\xi_{2b}^2)^\mathrm{inter}_{ij} &= \frac{
\sum_{n \mathbf{k}} \frac{g_{ij}^{n \mathbf{k}}}{(2\varepsilon_{n \mathbf{k}} - E_b)^2}
- \sum_{n, m \ne n, \mathbf{k}} \frac{ g_{ij}^{nm\mathbf{k}}  }
{(2\varepsilon_{n \mathbf{k}} - E_b)(2\varepsilon_{m \mathbf{k}} - E_b)} }
{\sum_{n \mathbf{k}} \frac{1}{(2\varepsilon_{n \mathbf{k}} - E_b)^2}},
\label{eqn:2binter}
\end{align}
offer a direct term-by-term comparison with the previous results on the 
inverse effective-mass tensor $(M_{2b}^{-1})_{ij}$ of the lowest-lying 
two-body bound states. For instance, the latter is also composed of an 
intraband contribution
$
(M_{2b}^{-1})^\mathrm{intra}_{ij} = \frac{2}{D} \sum_{\mathbf{k}} 
\partial_i \varepsilon_{n\mathbf{k}} \partial_j \varepsilon_{n\mathbf{k}}
/(2\varepsilon_{n \mathbf{k}} - E_b)^3
$
and an interband contribution
$
(M_{2b}^{-1})^\mathrm{inter}_{ij} = \frac{1}{D}
\sum_{n \mathbf{k}} g_{ij}^{n \mathbf{k}}/(2\varepsilon_{n \mathbf{k}} - E_b)
- \frac{1}{D}\sum_{n, m \ne n, \mathbf{k}} g_{ij}^{nm\mathbf{k}}/
(\varepsilon_{n \mathbf{k}} + \varepsilon_{m \mathbf{k}} - E_b),
$
where
$
D = \sum_{n \mathbf{k}} 1/(2\varepsilon_{n \mathbf{k}} - E_b)^2
$
~\cite{iskin24}.

In particular, we note that since the so-called geometric contribution 
$(\xi_{2b}^2)^\mathrm{inter}_{ij}$ is not due to the non-zero center-of-mass 
momentum $\mathbf{q}$ of the pair, its physical origin differs from that of 
$(M_{2b}^{-1})^\mathrm{inter}_{ij}$. In other words, while the interband 
contributions to the superfluid density, low-energy collective modes, 
GL coherence length, and similar quantities are all directly 
linked to each other through the effective-mass tensor of the Cooper 
pairs~\cite{iskin24, iskin24c}, and whose origin can be traced all the way back 
to $(M_{2b}^{-1})^\mathrm{inter}_{ij}$, the localization tensor is not 
one of them. On the other hand, despite their difference in origin, 
$(\xi_{2b}^2)^\mathrm{inter}_{ij}$ does not receive any contribution from 
band touchings, which is similar to $(M_{2b}^{-1})^\mathrm{inter}_{ij}$. 
Specifically, the first sum cancels the contributions from the second sum 
at the band touchings, i.e., whenever 
$\varepsilon_{n\mathbf{k}} = \varepsilon_{m\mathbf{k}}$ for any $n \neq m$.

It is also important to emphasize that Eqs.~(\ref{eqn:2bintra})
and~(\ref{eqn:2binter}) are exact for the lowest-lying 
$\mathbf{q} = \mathbf{0}$ bound state under the assumptions of time-reversal 
symmetry and uniform pairing, where Eq.~(\ref{eqn:2bintra}) is simply 
a sum over the well-known single-band expression~\cite{degennes66}.
In the case of an energetically-isolated flat band that is separated 
from the remaining bands with a finite band gap, they reduce to
\begin{align}
(\xi_{2b}^2)_{ij} \to \frac{1}{N_c} \sum_{\mathbf{k}} g_{ij}^{f \mathbf{k}}
\end{align}
in the $U/t \to 0$ limit, where $g_{ij}^{f \mathbf{k}}$ is the 
quantum metric of the flat band, which is in agreement with a 
recent preprint~\cite{ying24}. Note that
$
(M_{2b}^{-1})_{ij} \to \frac{U}{N} \sum_{\mathbf{k}} g_{ij}^{f \mathbf{k}}
$
in the very same limit~\cite{torma22}. Furthermore, we also note that 
$
(\xi_0^2)_{ij} \to \frac{1}{8 F N} \sum_{\mathbf{k}} g_{ij}^{f \mathbf{k}}
$
for the zero-temperature coherence length and 
$
(\xi_\mathrm{GL}^2)_{ij} \to \frac{1}{3 F N} \sum_{\mathbf{k}} g_{ij}^{f \mathbf{k}}
$
for the GL coherence length near the critical temperatures, i.e., for a 
dilute isolated flat-band superconductor when the particle filling 
$F \to 0$~\cite{iskin24c}.
Next, we benchmark Eqs.~(\ref{eqn:2bintra}) and~(\ref{eqn:2binter}) 
against an analogous tensor for the average size of Cooper pairs within 
the variational BCS mean-field theory, under the same assumptions.

\section{Mean-field BCS problem}
\label{sec:BCS}

Assuming time-reversal symmetry for the Bloch Hamiltonian and uniform 
pairing across the lattice sites, the BCS ground state can be written 
as~\cite{iskin24}
\begin{align}
|\mathrm{BCS} \rangle = \prod_{n \mathbf{k}}
\big( u_{n \mathbf{k}} + v_{n \mathbf{k}} c_{n \mathbf{k} \uparrow}^\dagger
c_{n, -\mathbf{k}, \downarrow}^\dagger \big)| 0 \rangle,
\label{eqn:BCS}
\end{align}
where 
$
u_{n \mathbf{k}} = \sqrt{\frac{1}{2} + \frac{\xi_{n\mathbf{k}}}{2E_{n\mathbf{k}}}}
$
and
$
v_{n \mathbf{k}} = \sqrt{\frac{1}{2} - \frac{\xi_{n\mathbf{k}}}{2E_{n\mathbf{k}}}}
$
are the usual coherence factors. Here, 
$
\xi_{n\mathbf{k}} = \varepsilon_{n\mathbf{k}} - \mu
$
is the energy measured relative to the chemical potential $\mu$, and 
$
E_{n\mathbf{k}} = \sqrt{\xi_{n\mathbf{k}}^2 + \Delta_0^2}
$
is the quasiparticle dispersion. The BCS expectation value 
$
\Delta_{S \mathrm{i}} = U \langle 
c_{S \mathrm{i} \uparrow} c_{S \mathrm{i} \downarrow}
\rangle
$
is taken as $\Delta_0$ for every sublattice site $S$ within any unit 
cell $\mathrm{i}$, representing the uniform BCS order parameter for 
pairing, i.e.,
$
\Delta_0 \equiv \frac{1}{N} \sum_{S \mathrm{i}} \Delta_{S \mathrm{i}}, 
$
leading to
$
\Delta_0 = \frac{U}{N} \sum_{n \mathbf{k}}
\langle c_{n \mathbf{k} \uparrow} c_{n,- \mathbf{k}, \downarrow} \rangle,
$ 
which is assumed to be real without loss of generality. 
In the zero-temperature BCS-BEC crossover 
formalism~\cite{pistolesi94, engelbrecht97}, it is sufficient to find 
$\mu$ and $\Delta_0$ from the self-consistent solutions of the mean-field 
gap equation
$
1 = \frac{U}{N} \sum_{n\mathbf{k}} 1/(2E_{n\mathbf{k}})
$
and the mean-field number equation
$
F = 1 - \frac{1}{N} \sum_{n\mathbf{k}} \xi_{n\mathbf{k}} / E_{n\mathbf{k}},
$
where the particle filling $0 \le F = \mathcal{N}/N \le 2$ corresponds to 
the total number of particles per lattice site.

Motivated by the pair-correlation function with opposite spins, 
the wave function for the Cooper pairs can be written as
~\cite{pistolesi94, engelbrecht97, schunck08, yu2014}
\begin{align}
\Phi(\mathsf{r}_1, \mathsf{r}_2) = \langle \mathrm{BCS}|
\psi_\uparrow^\dagger(\mathsf{r}_1) \psi_\downarrow^\dagger(\mathsf{r}_2)
| \mathrm{BCS} \rangle,
\label{eqn:Phi}
\end{align}
where the operator
$
\psi_\sigma^\dagger(\mathsf{r}) = \sum_{n \mathbf{k}} 
\phi_{n \mathbf{k} \sigma}^*(\mathsf{r}) c_{n \mathbf{k} \sigma}^\dagger
$
creates a spin-$\sigma$ fermion at position $\mathsf{r}$ and 
$
\phi_{n \mathbf{k} \sigma} (\mathsf{r})
$
is the Bloch wave function. Inserting
$
\langle \mathrm{BCS}|
c_{n \mathbf{k} \uparrow}^\dagger
c_{m \mathbf{k'} \downarrow}^\dagger
| \mathrm{BCS} \rangle = \delta_{nm} \delta_{\mathbf{k}, -\mathbf{k'}} 
u_{n\mathbf{k}} v_{n\mathbf{k}}
$
in Eq.~(\ref{eqn:Phi}), we find 
$
\Phi(\mathsf{r}_1, \mathsf{r}_2) = 
\frac{1}{N_c}
\sum_{n \mathbf{k}} e^{-\mathrm{i} \mathbf{k} \cdot \mathbf{r}} 
u_{n\mathbf{k}} v_{n\mathbf{k}}
n_{\mathbf{k} \uparrow}^* (\mathsf{r}_1)
n_{-\mathbf{k}, \downarrow}^* (\mathsf{r}_2).
$
Thus, the role of $\alpha_{nn \mathbf{k}}^\mathbf{0}$ in the two-body wave 
function $\Psi_\mathbf{0}(\mathsf{r}_1, \mathsf{r}_2)$ is effectively played by 
$
\frac{1}{\sqrt{A_{Cp}}} u_{n \mathbf{k}} v_{n \mathbf{k}}
$
in $\Phi(\mathsf{r}_1, \mathsf{r}_2)$, where 
$
A_{Cp} = \sum_{n \mathbf{k}} u_{n \mathbf{k}}^2 v_{n \mathbf{k}}^2 
= \sum_{n \mathbf{k}} \Delta_0^2/ (4 E_{n\mathbf{k}}^2)
$
is the normalization factor. Note that this observation is consistent 
with the number of condensed pairs, which is determined by
$
\sum_{n SS' \mathbf{k}}
|u_{n \mathbf{k}} v_{n \mathbf{k}} n_{S \mathbf{k}} n_{S' \mathbf{k}}^*|^2
 = A_{Cp},
$
i.e., the mean-field expression for the filling of condensed 
particles is
$
F_c = \frac{1}{N} \sum_{n\mathbf{k}} \Delta_0^2/(2E_{n\mathbf{k}}^2)
$
~\cite{iskin24, leggett}.

Analogous to the localization tensor introduced in Eq.~(\ref{eqn:loctensor}), 
the average size of Cooper pairs can be characterized using a related tensor
\begin{align}
(\xi_{Cp}^2)_{ij} = \sum_{\mathsf{r}_1 \mathsf{r}_2}
r_i r_j |\Phi (\mathsf{r}_1,\mathsf{r}_2)|^2,
\end{align}
where $\mathsf{r} = \mathsf{r}_1 - \mathsf{r}_2$ is the relative position 
between two particles, and $r_i$ is its $i$-component.
Having continuum systems in mind, the pair size is typically defined 
as the trace of the localization tensor in the BCS-BEC crossover 
theories~\cite{pistolesi94, engelbrecht97, schunck08, yu2014}. 
However, here we extend this concept to its tensorial form to highlight 
its deep connection to the quantum-metric tensor in multiband Hubbard models.
Following a similar approach as in Eq.~(\ref{eqn:xi_ij}), we find that
$
(\xi_{Cp}^2)_{ij} = \frac{1}{A_{Cp}} \sum_{nmSS' \mathbf{k}}
\partial_i \big( u_{n \mathbf{k}} v_{n \mathbf{k}}
n_{S \mathbf{k}} n_{S' \mathbf{k}}^* \big)
\partial_j \big( u_{m \mathbf{k}} v_{m \mathbf{k}}
m_{S \mathbf{k}}^* m_{S' \mathbf{k}} \big).
$
This expression can also be separated into intraband and interband 
contributions as
$
(\xi_{Cp}^2)_{ij} = (\xi_{Cp}^2)^\mathrm{intra}_{ij} 
+ (\xi_{Cp}^2)^\mathrm{inter}_{ij},
$
where
\begin{align}
\label{eqn:mbintra}
(\xi_{Cp}^2)^\mathrm{intra}_{ij} &= \frac{
\sum_{n \mathbf{k}} 
\partial_i \varepsilon_{n\mathbf{k}}
\partial_j \varepsilon_{n\mathbf{k}} 
\frac{\xi_{n \mathbf{k}}^2}{E_{n \mathbf{k}}^6}}
{\sum_{n \mathbf{k}} \frac{1}{E_{n \mathbf{k}}^2}}, \\
(\xi_{Cp}^2)^\mathrm{inter}_{ij} &= \frac{
\sum_{n \mathbf{k}} \frac{g_{ij}^{n \mathbf{k}}} {E_{n \mathbf{k}}^2}
- \sum_{n, m \ne n, \mathbf{k}}
\frac{g_{ij}^{nm\mathbf{k}}} {E_{n \mathbf{k}} E_{m \mathbf{k}}}}
{\sum_{n \mathbf{k}} \frac{1}{E_{n \mathbf{k}}^2}}.
\label{eqn:mbinter}
\end{align}
Note that Eq.~(\ref{eqn:mbintra}) is simply a sum over the well-known 
single-band expression~\cite{pistolesi94, engelbrecht97}. 
In the context of cold Fermi gases, this latter length scale was 
shown to qualitatively agree with the pair size extracted from 
radio-frequency-spectrum measurements across the BCS-BEC 
crossover~\cite{schunck08}.
Furthermore, similarly to $(\xi_{2b}^2)^\mathrm{inter}_{ij}$, 
$(\xi_{Cp}^2)^\mathrm{inter}_{ij}$ also receives no contribution from 
band touchings, i.e., the first sum cancels those touching 
contributions from the second sum whenever 
$
\varepsilon_{n\mathbf{k}} = \varepsilon_{m\mathbf{k}}
$ 
for any $n \neq m$.
It is also reassuring to see that Eqs.~(\ref{eqn:mbintra}) and~(\ref{eqn:mbinter})
reproduce Eqs.~(\ref{eqn:2bintra}) and~(\ref{eqn:2binter}), respectively, 
in the dilute limit $F \ll 1$ where $\mu \to E_b/2 < 0$ and 
$\xi_{n \mathbf{k}} \gg \Delta_0$ for every state in the BZ. 
This is typically the case in the BEC limit when $U/t \gg 1$, but not when
moving away from it towards the BCS regime of dispersive bands. 

The pair size is a meaningful observable, representing the characteristic 
length scale associated with the pair-correlation function. Its physical 
significance is well-established in the BCS-BEC crossover literature~\cite{pistolesi94, engelbrecht97, schunck08, yu2014}. 
For instance, in the continuum limit, the pair size aligns (up to a factor 
of order unity) with the phase coherence length in the BCS regime, 
while it precisely matches the size of two-body bound states in the 
BEC regime. This duality underscores its utility in capturing pair 
correlations across the crossover.

\section{Numerical Illustration}
\label{sec:numerics}

As a numerical illustration of our analytical expressions, we study the 
pyrochlore lattice, which is obtained by constructing the line graph of the 
diamond lattice, featuring two degenerate flat bands in three dimensions. 
Its crystal structure is a face-centered-cubic Bravais lattice with a 
four-point basis, resulting in a truncated-octahedron-shaped BZ. 
Recent demonstrations of flat bands and superconductivity in materials 
such as the pyrochlore metal CaNi$_2$~\cite{wakefiel23} and the pyrochlore 
superconductor CeRu$_2$~\cite{huang23} highlight the growing relevance of 
this model in understanding emergent quantum phenomena.
Its Bloch Hamiltonian is determined by
$
h_{SS \mathbf{k}}^\sigma = 0,
$
$
h_{AB \mathbf{k}}^\sigma = -2\bar{t} \cos\big(\frac{k_y+k_z}{4}a\big),
$
$
h_{AC \mathbf{k}}^\sigma = -2\bar{t} \cos\big(\frac{k_x+k_z}{4}a\big),
$
$
h_{AD \mathbf{k}}^\sigma = -2\bar{t} \cos\big(\frac{k_x+k_y}{4}a\big),
$
$
h_{BC \mathbf{k}}^\sigma = -2\bar{t} \cos\big(\frac{k_x-k_y}{4}a\big),
$
$
h_{BD \mathbf{k}}^\sigma = -2\bar{t} \cos\big(\frac{k_x-k_z}{4}a\big)
$
and
$
h_{CD \mathbf{k}}^\sigma = -2\bar{t} \cos\big(\frac{k_y-k_z}{4}a\big),
$
where $\bar{t}$ is the tight-binding hopping parameter between 
nearest-neighbor sites and $a$ is the side-length of the conventional 
simple-cubic cell~\cite{iskin24, iskin24c}. 
The resulting Bloch spectrum comprises two 
dispersive bands given by
$
\varepsilon_{1\mathbf{k}\sigma} = -2\bar{t}(1 + \sqrt{1 + \gamma_\mathbf{k}})
$
and
$
\varepsilon_{2\mathbf{k}\sigma} = -2\bar{t}(1 - \sqrt{1 + \gamma_\mathbf{k}}),
$
where 
$
\gamma_\mathbf{k} = \cos(k_x a/2) \cos(k_y a/2) + 
\cos(k_y a/2) \cos(k_z a/2) + \cos(k_x a/2) \cos(k_z a/2),
$
as well as two degenerate flat bands given by
$
\varepsilon_{3\mathbf{k}\sigma} = \varepsilon_{4\mathbf{k}\sigma} = 2\bar{t}.
$
To position these flat bands at the bottom of the spectrum, we set 
$\bar{t} \to -t$ and choose $t > 0$ as the unit of energy. Note that 
$\varepsilon_{2\mathbf{k}\sigma}$ touches the flat bands at $\mathbf{k = 0}$. 

In our previous work on the pyrochlore lattice~\cite{iskin24}, 
we explored various connections between the superfluid-weight tensor 
and the effective-mass tensor of the lowest-lying two-body branch at 
zero temperature, the kinetic coefficient in the GL theory 
near the critical temperature, and the velocity of low-energy Goldstone 
modes at zero temperature~\cite{iskin24}. In addition, we analyzed 
the GL coherence length near the critical temperature and 
compared it with the zero-temperature coherence length, both of which 
are tied to the effective-mass tensor of Cooper pairs~\cite{iskin24c}. 
While these studies also focus on the pyrochlore lattice, the topics 
they address are distinct from the current paper, with no overlap.

\begin{figure} [htb]
\includegraphics[width = 0.99\linewidth]{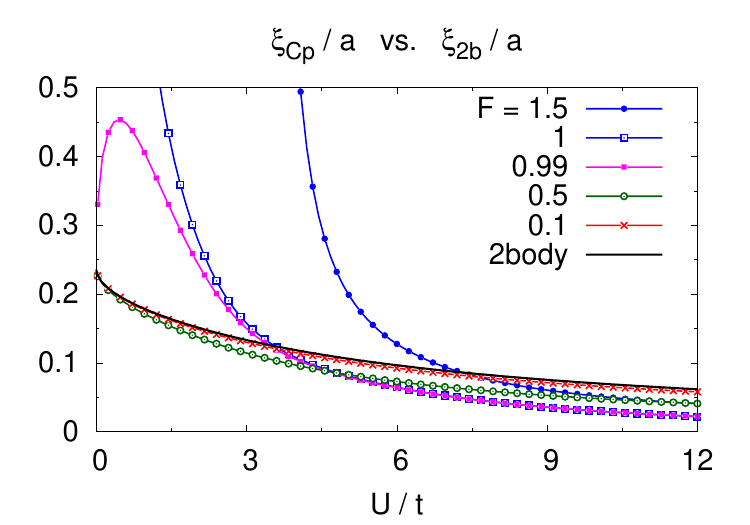}
\caption{\label{fig:xi}
While the average size of Cooper pairs diverges in the BCS regime 
$1 < F < 2$ of dispersive bands as $\Delta_0/t \to 0$ in the $U/t \to 0$ 
limit, it approaches the size of the lowest-lying two-body bound state 
in the dilute flat-band regime for any $U \ne 0$, as the particle 
filling $F \to 0$.
}
\end{figure}

Since the pyrochlore-Hubbard model exhibits both time-reversal symmetry 
and uniform-pairing condition~\cite{iskin24, iskin24c}, we can directly 
apply Eqs.~(\ref{eqn:2bintra}) and~(\ref{eqn:2binter}) to the two-body 
problem and Eqs.~(\ref{eqn:mbintra}) and~(\ref{eqn:mbinter}) to the 
many-body problem. The results are shown in Fig.~\ref{fig:xi}, where 
$
(\xi_{2b}^2)_{ij} = \xi_{2b}^2 \delta_{ij} 
$
and
$
(\xi_{Cp}^2)_{ij} = \xi_{Cp}^2 \delta_{ij} 
$
as a consequence of uniform pairing. In the absence of interactions, when 
$U = 0$, $-2t < \mu < 6t$ lies within the dispersive bands when $1 < F < 2$. 
Here, $\mu = 2t$ corresponds to $F = 1.5$ and $\mu \to -2t$ from above 
corresponds to a half-filled lattice with $F \to 1$ from above. 
Note that $\mu = -2t$ coincides with the degenerate flat bands when $0 < F < 1$. 
Thus, Fig.~\ref{fig:xi} shows that the pair size $\xi_{Cp}$ diverges 
in the BCS regime $1 < F < 2$ of dispersive bands as $\Delta_0/t \to 0$ 
in the $U/t \to 0$ limit, 
which is consistent with the BCS behavior in the usual BCS-BEC 
crossover problem~\cite{pistolesi94, engelbrecht97}. 
In contrast, $\xi_{Cp}$ remains finite and relatively small for the 
flat-band regime $0 < F < 1$ as $\Delta_0/t \to 0$ in the $U/t \to 0$ 
limit, and does not diverge.
The striking difference between the $U/t \to 0$ limit of the BCS regime 
of dispersive bands and the flat-band regime is in accordance with 
the modern theory of insulating states: 
the localization tensor diverges in the thermodynamic 
limit in any metal, while it remains finite in any insulator~\cite{resta11}. 
Due to the presence of compact localized states in the non-interacting 
spectrum, the flat-band regime behaves similarly to an insulating state.
Furthermore, Fig.~\ref{fig:xi} shows that $\xi_{Cp}$ approaches the 
size $\xi_{2b}$ of the lowest-lying two-body bound state for all 
$U \ne 0$ in the dilute limit as $F \to 0$. In this case, it can 
be shown that $\xi_{Cp} \to \xi_{2b}$ term by term for all $U \ne 0$. 
In particular, when $U/t \to 0$, we find
$
\xi_{2b}^\mathrm{intra} \to 0
$
and
$
(\xi_{2b}^\mathrm{inter})^2 \to 
\frac{1}{N_c} \sum_{m \notin f, \mathbf{k}}^{'} g_{xx}^{f m \mathbf{k}} 
\approx 0.056a^2,
$
where $f = \{3,4\}$ refers to the flat bands and $m = \{1,2\}$ refers to 
the dispersive bands, and the prime sum excludes the band touchings.
Thus, we expect $\xi_{2b} \approx 0.24a$ in the $U/t \to 0$ limit, 
which is consistent with Fig.~\ref{fig:xi}. Note that the shortest 
distance between lattice sites is approximately $a/\sqrt{8} \approx 0.35a$ 
for the pyrochlore lattice.

\begin{figure} [htb]
\includegraphics[width = 0.99\linewidth]{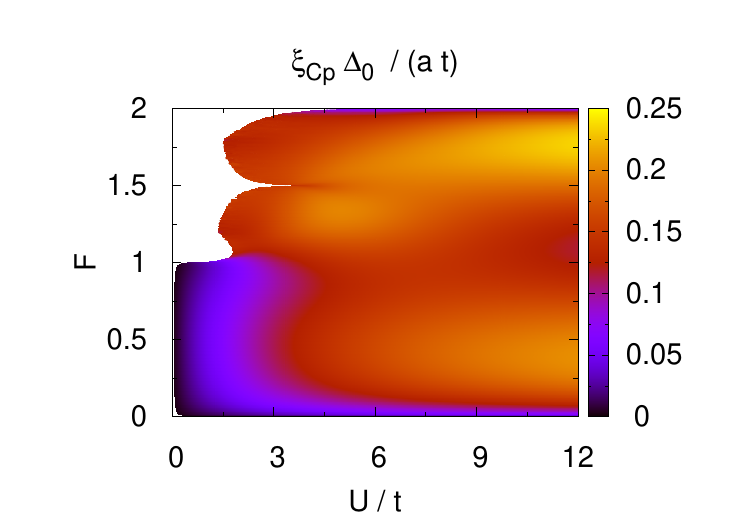}
\caption{\label{fig:ximap}
The average size of Cooper pairs scales as $t/\Delta_0$ in most of the 
parameter space, except for the case of flat-band superconductivity 
in the low-$U/t$ regime. Since our numerics become unreliable in the 
limit $\Delta_0/t \to 0$, we present only the data with $\Delta_0/t > 0.01$, 
which reveals the underlying single-particle density of states at the 
periphery of the white region, i.e., in the BCS regime of dispersive bands 
when $1 < F < 2$~\cite{iskin24}.
} 
\end{figure}

In Fig.~\ref{fig:ximap}, we present a color map of $\xi_{Cp}$ as a function 
of $F$ and $U/t$. This map shows that $\xi_{Cp}$ scales as $t/\Delta_0$ 
over most of the parameter space, which is in perfect agreement with 
the BCS result, except in the flat-band regime at low $U/t$. In the latter 
regime, $\xi_{Cp}$ is governed solely by the quantum geometry of the 
Bloch states. A similar scaling behavior has recently been reported for 
a related correlation length in the context of the sawtooth 
lattice~\cite{thumin24}. Furthermore, Fig.~\ref{fig:ximap} bears 
resemblance to results on zero-temperature coherence length, which also 
scales as $t/\Delta_0$ in most of the parameter space~\cite{iskin24c}. 
However, while coherence length and $\xi_{Cp}$ differ only by a factor of 
order unity around $F = 0.5$, i.e., around the half-filled flat-band regime, 
coherence length becomes much larger than $\xi_{Cp} \to \xi_{2b}$ 
in the dilute flat-band regime as $F \to 0$. 
For instance, the coherence length
$
\xi_0^2 \to \frac{1}{4 F N} \sum_{m \notin f, \mathbf{k}}^{'} 
g_{xx}^{f m \mathbf{k}} 
$
diverges as $F \to 0$ in the $U/t \to 0$ limit~\cite{iskin24c}.
This finding is consistent with the BEC behavior in the usual BCS-BEC crossover
problem~\cite{pistolesi96, engelbrecht97}.

\section{Conclusion}
\label{sec:conc}

In summary, by considering a multiband Hubbard model that exhibits 
time-reversal symmetry and uniform pairing in the lattice, we analyzed 
how the quantum geometry of the Bloch states affects: (i) the size of 
lowest-lying two-body bound state in vacuum through an exact calculation 
of the localization tensor, and (ii) the average size of Cooper pairs 
through a related tensor within the mean-field BCS-BEC crossover theory 
at zero temperature. Our primary finding is that, in contrast to the BCS 
regime of dispersive bands, where the pair size is known to diverge 
as $t/\Delta_0$ when the order parameter $\Delta_0/t \to 0$ in the 
weakly-interacting $U/t \to 0$ limit, 
it remains finite and relatively small in the 
flat-band regime under the very same conditions, perfectly matching 
the exact two-body result in the dilute limit.

The pair size bears resemblance to recent results on the zero-temperature 
coherence length, which also scales as $t/\Delta_0$ in most of the 
parameter space~\cite{iskin24c}. However, in the flat-band regime, 
the pair size remains finite and relatively small, 
being governed solely by the quantum metric 
in the dilute limit. Thus, we emphasized that the pair size and coherence 
length are distinct physical quantities, particularly in the dilute limit.
Furthermore, as revealed in this paper through an exact calculation of 
the two-body problem, their quantum-geometric origins in a multiband Hubbard 
model are also distinct. Similar to superfluid density, low-energy collective 
modes, and the GL coherence length, while the zero-temperature 
coherence length is directly related to the motion of Cooper pairs through 
the inverse effective-mass tensor~\cite{iskin24, iskin24c}, the pair size 
is not. In other words, the quantum-geometric origin of the pair size 
does not stem from the non-zero center-of-mass momentum of the pair, 
which distinguishes it from previous findings in this context. Thus, the 
quantum-geometric effects that influence these quantities operate in 
fundamentally different ways.

\begin{acknowledgments}
The author acknowledges funding from US Air Force Office of Scientific 
Research (AFOSR) Grant No. FA8655-24-1-7391.
\end{acknowledgments}

\appendix

\section{An unphysical length scale} 
\label{sec:appA}

Motivated by recent literature on spin-orbit-coupled Fermi gases~\cite{yu2014}, 
we propose an alternative, yet ultimately unsuccessful, approach to the 
wave function of Cooper pairs.
For this purpose, the BCS ground state, given by Eq.~(\ref{eqn:BCS}), 
can equivalently be written as
$
|\mathrm{BCS} \rangle = ( \prod_{n \mathbf{k}} u_{n \mathbf{k}} )  
e^{\sum_{n \mathbf{k}} \frac{v_{n \mathbf{k}}}{u_{n \mathbf{k}}} 
c_{n \mathbf{k} \uparrow}^\dagger c_{n, -\mathbf{k}, \downarrow}^\dagger} 
| 0 \rangle.
$
This reformulation suggests that the state
\begin{align}
|\Phi_p \rangle = \frac{1}{\sqrt{A_{p}}}\sum_{n \mathbf{k}}  
\frac{v_{n \mathbf{k}}}{u_{n \mathbf{k}}} 
c_{n \mathbf{k} \uparrow}^\dagger c_{n, -\mathbf{k}, \downarrow}^\dagger 
| 0 \rangle
\end{align}
may represent Cooper pairs in a many-body setting, where 
$
A_{p} = \sum_{n \mathbf{k}} v_{n \mathbf{k}}^2 / u_{n \mathbf{k}}^2 
= \sum_{n \mathbf{k}}(E_{n\mathbf{k}} - \xi_{n\mathbf{k}})^2/\Delta_0^2
$
is the normalization factor. Thus, the role of $\alpha_{nn \mathbf{k}}^\mathbf{0}$ 
in the two-body ansatz $|\Psi_\mathbf{0} \rangle$ is effectively played 
by 
$
\frac{1}{\sqrt{A_{p}}} \frac{v_{n \mathbf{k}}}{u_{n \mathbf{k}}}
$
in $|\Phi_p \rangle$. Then, following a similar approach as in 
Eq.~(\ref{eqn:xi_ij}), we find that 
$
(\xi_p^2)_{ij} = \frac{1}{A_{p}} \sum_{nmSS' \mathbf{k}}
\partial_i \big( \frac{v_{n \mathbf{k}}}{u_{n \mathbf{k}}}
n_{S \mathbf{k}} n_{S' \mathbf{k}}^* \big)
\partial_j \big( \frac{v_{m \mathbf{k}}}{u_{m \mathbf{k}}}
m_{S \mathbf{k}}^* m_{S' \mathbf{k}} \big),
$
and separate it into intraband and interband contributions as
$
(\xi_{p}^2)_{ij} = (\xi_{p}^2)^\mathrm{intra}_{ij} 
+ (\xi_{p}^2)^\mathrm{inter}_{ij},
$
where
\begin{align}
\label{eqn:pintra}
(\xi_p^2)^\mathrm{intra}_{ij} &= \frac{
\sum_{n \mathbf{k}} \partial_i \varepsilon_{n\mathbf{k}}
\partial_j \varepsilon_{n\mathbf{k}} 
\Big( 1 - \frac{\xi_{n \mathbf{k}}}{E_{n \mathbf{k}}} \Big)^2}
{\sum_{n \mathbf{k}} (E_{n \mathbf{k}} - \xi_{n \mathbf{k}})^2}, \\
(\xi_p^2)^\mathrm{inter}_{ij} &= \frac{
\sum_{n \mathbf{k}} (E_{n \mathbf{k}} - \xi_{n \mathbf{k}})^2 g_{ij}^{n \mathbf{k}}}
{{\sum_{n \mathbf{k}} (E_{n \mathbf{k}} - \xi_{n \mathbf{k}})^2 }}
\nonumber \\
&- \frac{
\sum_{n, m \ne n, \mathbf{k}}
(E_{n \mathbf{k}} - \xi_{n \mathbf{k}})(E_{m \mathbf{k}} - \xi_{m \mathbf{k}})
g_{ij}^{nm\mathbf{k}} }
{{\sum_{n \mathbf{k}} (E_{n \mathbf{k}} - \xi_{n \mathbf{k}})^2 }}.
\label{eqn:pinter}
\end{align}
Note that, similarly to $(\xi_{2b}^2)^\mathrm{inter}_{ij}$ and 
$(\xi_{Cp}^2)^\mathrm{inter}_{ij}$, $(\xi_{p}^2)^\mathrm{inter}_{ij}$ also 
receives no contribution from band touchings.

\begin{figure} [htb]
\includegraphics[width = 0.99\linewidth]{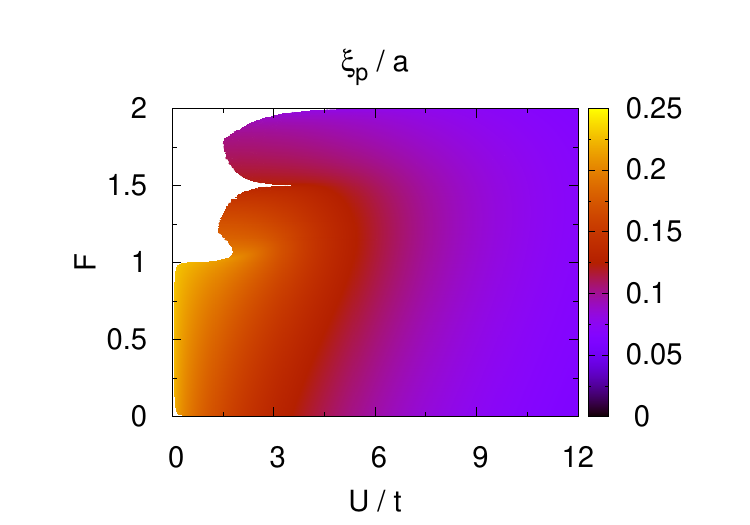}
\caption{\label{fig:app}
In the low-$U/t$ regime, while the length scale $\xi_p$ qualitatively 
reproduces the correct results for the average size of Cooper pairs 
in the flat-band regime, e.g., it coincides perfectly with the size of the 
lowest-lying two-body bound state in the dilute flat-band limit as $F \to 0$, 
it gives unphysical results for the BCS regime of dispersive bands when 
$1 < F < 2$, as it saturates in that range even though $\Delta_0/t \to 0$.
} 
\end{figure}

In Fig.~\ref{fig:app}, we present a color map of $\xi_p$ for the 
pyrochlore lattice, demonstrating that $\xi_p$ does not diverge in the BCS
regime $U/t \to 0$, even as $\Delta_0/t \to 0$ for the dispersive bands
when $1 < F < 2$. Since this behavior contrasts sharply with physical
expectations~\cite{pistolesi94, engelbrecht97}, we conclude that $\xi_p$ does not 
represent a physically meaningful observable length scale. On the other hand, 
$\xi_p$ coincides perfectly with $\xi_{2b}$ and $\xi_{Cp}$ for any 
$U \ne 0$ in the dilute flat-band limit when $F \to 0$. In this particular 
case, assuming 
$
\xi_{n \mathbf{k}} \gg \Delta_0
$ 
for every state in the BZ, it can be shown analytically 
that Eqs.~(\ref{eqn:pintra}) and~(\ref{eqn:pinter}) are identical to 
Eqs.~(\ref{eqn:mbintra}) and~(\ref{eqn:mbinter}), respectively,
where
$
v_{n \mathbf{k}} / u_{n \mathbf{k}} \approx 
v_{n \mathbf{k}} u_{n \mathbf{k}} \approx
\Delta_0 / (2 \xi_{n \mathbf{k}})
$
at the leading order.

\bibliography{refs}

\end{document}